\documentstyle[aps,prb,epsfig,preprint,amstex]{revtex}
\input epsf

\begin{document}

\draft
\tightenlines

%aliased shortcut commands
\newcommand{\usq}{$\langle u^2\rangle$ }
\newcommand{\deltarsq}{$\langle \delta r^2\rangle$ }
\newcommand{\usqn}{$\langle u^2\rangle_N$ }
\newcommand{\usqinf}{$\langle u^2\rangle_\infty$ }
\newcommand{\he}{$^4$He }

\title{The Debye-Waller factor in solid $^3$He and $^4$He}
\author{E. W. Draeger and D. M. Ceperley}
\address{Department of Physics and National Center for Supercomputing Applications,
University of Illinois, Urbana-Champaign, 61801}
\maketitle

\begin{abstract}
The Debye-Waller factor and the mean-squared displacement from lattice
sites for solid $^3$He and \he were calculated with Path Integral
Monte Carlo at temperatures between 5~K and 35~K, and densities
between 38~nm$^{-3}$ and 67~nm$^{-3}$.  It was found that the
mean-squared displacement exhibits finite-size scaling consistent with
a crossover between the quantum and classical limits of $N^{-2/3}$ and
$N^{-1/3}$, respectively.  The temperature dependence appears to be
$T^3$, different than expected from harmonic theory.  An anisotropic
$k^4$ term was also observed in the Debye-Waller factor, indicating
the presence of non-Gaussian corrections to the density distribution
around lattice sites.  Our results, extrapolated to the thermodynamic
limit, agree well with recent values from scattering experiments.

\end{abstract}

\section{Introduction}

Solid helium at low temperatures is the prototype quantum
crystal\cite{glyde}, since its ground state is not well-described by
harmonic perturbations about a minimum potential configuration.  For
temperatures between 5~K and 35~K, and densities between 38~nm$^{-3}$
and 67~nm$^{-3}$, the kinetic energy is always greater than about 25 K
and dominates over thermal energies.  Solid helium has a wide range of
experimentally accessible temperatures and densities.  With new
scattering sources, the density distribution and other correlation
functions can be measured with unprecedented accuracy, allowing for a
careful comparison between theory and experiment.

It is possible to calculate the properties of helium very accurately
using Monte Carlo methods, because the inter-atomic potential
is accurately known, more accurately than any other atomic or
molecular condensed matter system.  Additionally, for bosonic and
distinguishable particle systems, Path Integral Monte Carlo methods
can calculate equilibrium properties directly from an assumed
Hamiltonian without significant approximation. Even in the case of
solid $^3$He, effects of Fermi statistics can be neglected for
temperatures above 0.1 K and densities slightly away from melting. 

In scattering, the Debye-Waller factor is the fractional intensity
shift due to recoilless processes, and can be directly related to the
one-dimensional mean-squared displacement of particles from their
lattice sites $\langle u^2\rangle$.  In this paper we compute \usq and
compare it to experimental results, obtaining good agreement.  In the
process we make several interesting observations.  First, the density
distribution of helium atoms is slightly non-Gaussian, a fact that had
been speculated on but not yet observed.  Second, we observe some
unusual dependence of \usq on the size of the finite system being
simulated, indicating a scaling crossover between the quantum and
classical regimes.  The dependence of \usq on the number of atoms $N$
being simulated appears to be well-described by harmonic theory.
Finally, the temperature dependence appears to be $T^3$, rather than
the $T^2$ predicted by harmonic theory.

\subsection{The Debye-Waller Factor}
\label{dwtheory}
The static structure factor, as measured in scattering, is defined
as
\begin{equation}
S(k) = \langle\left|\rho_k\right|^2\rangle
\label{sk1}
\end{equation}
where
$\rho_k\equiv\frac{1}{\sqrt{N}}\sum\limits_ie^{i\bf{k}\cdot\bf{r}_i}$.
The structure factor can be computed directly with PIMC for finite
systems of $N \lesssim 10000$ atoms. In a solid, the structure factor
has large peaks at the reciprocal lattice vectors of the perfect
lattice. It is usually assumed that the magnitude of $S(k)$ behaves as
\begin{equation}
S(k) \propto \exp ( -2W ) \cong \exp ( -k^2\langle u^2\rangle )
\label{harmdw}
\end{equation}
where $\exp(-2W)$ is the Debye-Waller factor.  This equation relies on
the assumption that the particle densities are normally distributed
about the lattice sites.

We will now derive a more general form of Eq.~(\ref{harmdw}).  We will
assume a finite system, with periodic boundary conditions and
precisely $N$ particles at positions ${\bf r}_i$ and $N$ perfect
lattice sites ${\bf Z}_i$.  Using particle symmetry, we can rewrite
Eq.~(\ref{sk1}) as:
\begin{equation}
S(k) = 1 + (N-1)\langle e^{i\bf{k}\cdot
(\bf{r}_1-\bf{r}_j)}\rangle_j,
\label{sk2}
\end{equation}
where the angle brackets $\langle \rangle_j$ denote an average over
both the thermal density matrix {\it and} particles $j\neq1$. When
$\bf{k}$ is a reciprocal lattice vector,
\begin{equation}
e^{i\bf{k}\cdot(\bf{r}_1-\bf{r}_j)} = e^{ik(u_1-u_j)},
\end{equation}
where $u_i=\hat{\bf k}\cdot ( \bf{r}_i-\bf{Z}_i)$ is the displacement
of particle i from its lattice site in the direction of $\bf{k}$.  We
assume a simple Bravais lattice, and assign each particle to a lattice
site.  Now consider the variable $x \equiv u_1-u_j$. Using the
cumulant expansion\cite{vankampen} to evaluate the average of an
exponential in terms of the moments of $x$, we can write:
\begin{align}
\langle e^{ikx}\rangle = \exp \left(  - \frac{k^2}{2} \langle
x^2\rangle   + \frac{k^4}{24}\left( \langle x^4\rangle  - 3\langle
x^2\rangle^2 \right) + \ldots \right),
\label{sk3}
\end{align}
since the odd powers of $x$ in the expansion vanish under the
interchange $1\leftrightarrow j$ allowed by particle symmetry.  For a
system much larger than the correlation length $\xi$ of $u$, which is
finite in solid helium, $u_1$ is uncorrelated with $u_j$ except for
the neighbors of particle $1$.  Hence:
\begin{align}
&\langle x^2\rangle =2\langle u^2\rangle-2\langle u_1 u_j\rangle
\cong 2\langle u^2\rangle + {\cal O}\left(\left(\xi/L\right)^3\right)\nonumber \\ &\langle x^4\rangle =
2\langle u^4\rangle + 6\langle u^2\rangle^2 -4\langle u_1^3u_j\rangle -
4\langle u_1u_j^3\rangle \cong
2\langle u^4\rangle + 6\langle u^2\rangle^2 + {\cal O}\left(\left(\xi/L\right)^3\right).
\label{sk4}
\end{align}
Here, $L$ is the box length.  Combining
Eqs. (\ref{sk2},\ref{sk3},\ref{sk4}), we obtain
\begin{equation}
S(k) \cong 1 + (N-1)\exp\left( -k^2\langle u^2\rangle + \frac{\kappa
k^4}{12}\langle u^2\rangle^2 \right),
\label{sk5}
\end{equation}
where the kurtosis $\kappa$ is defined as the relative deviation of
the fourth moment from a normal distribution,
\begin{equation}
\kappa \equiv \frac{\langle u^4 \rangle}{\langle u^2\rangle^2} - 3.
\label{kurt1}
\end{equation}
It vanishes if the density distribution is normal in the scattering
direction.
Previous analysis of this kind has assumed Gaussian fluctuations, and
hence neglected all higher order terms beyond the first. However, we
find that the kurtosis is not precisely zero and has directional
dependence.

\section{PIMC calculations of solid helium}

Path Integral Monte Carlo simulations were performed as discussed in
ref. \cite{ceperley}.  The system being simulated consists of
particles (Boltzmannons) in a box with periodic boundary conditions at
a fixed density. The helium atoms were assumed to interact pair-wise
with the Aziz potential\cite{aziz}. Although small errors in energy
are expected with this potential due to the absence of three-body
interactions, we expect the pair potential to describe well the
density distribution due to the fact that three-body and higher order
contributions are smoothly varying with respect to atomic positions.
We implicitly test this assumption by comparing to experimental
values. The pair potential was set to zero for inter-atomic distances
greater than 6~\AA. We determined that this is the minimum cut-off
that could be used without causing systematic errors in the
mean-squared displacement. The pair potential was used to compute the
exact pair density matrix for the system with an imaginary time-step
equal to $\tau=1/160$~K$^{-1}$. Using this action, the time-step error
was found to be negligible. We used neighbor lists to achieve linear
scaling of the computer time versus the number of atoms, and were able
to simulate systems of up to 3000 atoms.

We computed the Debye-Waller factor two different ways.  First,
\usq was computed directly from the distance of the atoms from
their lattice sites,
\begin{equation}
\langle u^2 \rangle = \frac{1}{3}\left\langle
\frac{1}{NM}\sum_{i=1}^N\sum_{j=1}^{M}({\bf r}_{ij}-{\bf Z}_i)^2\right\rangle
\label{usq1}
\end{equation}
where $M$ is the number of imaginary time slices, $N$ is the number of
particles, and ${\bf r}_{ij}$ is the position of particle $i$ at
imaginary time slice $j$.  The factor of 1/3 arises because in a cubic
crystal we can also average over the three spatial directions.

For the sake of convenience, it is useful to forbid particle exchanges
between lattice sites so that the particles do not have to be
periodically re-assigned to the nearest site.  We assured localization
to a lattice site by ``tethering'' each particle: specifying a
distance from its lattice site past which all attempted moves were
rejected. The tether distance, 2.6~\AA, was chosen to be on the order
of, but slightly less than, the average nearest-neighbor distance and
did not introduce a noticeable change in $\langle u^2\rangle$ or the
structure factor.

The second method for computing the Debye-Waller factor was to
calculate $S(k)$ directly from Eq.~(\ref{sk1}) from the set of all
reciprocal lattice vectors of $k \leq 9$~\AA$^{-1}$ and then use
Eq.~(\ref{sk5}) to determine \usq and the kurtosis $\kappa$ by a
least-squares fit to $\log((S(k)-1)/(N-1))$.  This method has the
advantage of not requiring tethering or indeed any of the assumptions
used in deriving Eq.~(\ref{sk5}).  It also allows one to determine not
only \usq but also any non-Gaussian components.  Agreement between the
two approaches shows that correlations in $u$ in Eq.~(\ref{sk4}) do
decay rapidly. Calculating \usq either directly with Eq.~ (\ref{usq1})
or from fitting to $S(k)$ always gave the same value within
statistical error, with similar error bars (see Fig.~[\ref{bdw50}]).

\section{Finite-Size Scaling}

Before we can present the comparison to experiment, we must
extrapolate the results obtained for a finite system to the
thermodynamic limit. We observed a very slow convergence. To carefully
examine the finite-size effects we simulated much larger systems (up
to 3000 atoms) than had been done previously with PIMC.  Even with the
increased range of system size, the finite-size effects are not
well-described by a power law, but are instead in a crossover region.
For the temperature and density values studied in this paper, the
crossover region appears to span the range of system sizes available
to present computer simulation ($N_{\rm crossover} \le 10^5$),
requiring careful fitting to obtain values in the thermodynamic limit.

Young and Alder\cite{young} have used Debye theory to analyze the
finite-size dependence of a system of classical hard spheres and
determined that $\langle u^2\rangle \propto \rho^{-2/3} T N ^{-1/3}$.
They found that this scaling was accurately able to fit values
obtained with molecular dynamics simulations. Recent classical Monte
Carlo calculation on a Lennard-Jones model of an fcc solid
\cite{shulka} also found the same dependence.

Runge and Chester \cite{runge} looked at the size effects of a hard
sphere system using PIMC. Using the same Debye theory but now taking
quantum effects into account, they estimated that the finite-size
effects scale as $N^{-2/3}$ at zero temperature, and found a crossover
from the quantum to the classical scaling ($N^{-1/3}$) as the
temperature was increased.

We used harmonic theory to derive a reasonable functional form for the
finite-size effects of $\langle u^2\rangle$ in a crossover region and
the width of that crossover region.  In general, one can write the
mean-squared displacement as\cite{mahan}
\begin{equation}
\langle u^2 \rangle = a^3\int \frac{d^3{\bf k}}{(2\pi)^3} \int
\frac{d\omega}{2\pi}\frac{A({\bf k},\omega)}{e^{\beta\hbar\omega}-1}
\label{displace}
\end{equation}
where $A({\bf k},\omega)$ is the spectral function for the
displacement-displacement Green's function, $a^3 = 1/\rho$ is the
volume of the unit cell, and $\beta = 1/k_BT$.  It is useful to define
\begin{equation}
A({\bf k}) = \int\limits_{-\infty}^{\infty} \frac{d\omega}{2\pi}\frac{A({\bf k},\omega)}{e^{\beta\hbar\omega}-1}
\end{equation}
so that 
\begin{equation}
\langle u^2\rangle = \frac{a^3}{(2\pi)^3}\int d^3{\bf k}\;A({\bf k}).
\end{equation}

Let us assume that the effect of the periodic boundary conditions is
to replace the integral in Eq.~(\ref{displace}) with a sum:  
\begin{equation}
\delta\langle u^2\rangle \equiv \langle u^2\rangle_{\infty}-\langle
u^2\rangle_N =\frac{a^3}{(2\pi)^3}\int d^3{\bf k}\;A({\bf k}) - \frac{a^3}{(2\pi)^3}\sum_{{\bf k}\neq0}
A({\bf k})\;k_c^3.
\label{finerr}
\end{equation} 
The finite spacing of ${\bf k}$ is a function of the system size, and
is given by $k_c = 2\pi(\rho/N)^{1/3}$ for a cubic simulation cell.
The main contribution to the finite-size error is due to the omission
of the ${\bf k}=0$ value.  If we further assume that for small values
of $k$ that $A({\bf k})$ factors into an analytic function (smooth and
continuous near $k=0$) and a singular factor $|k|^{-\nu}$, then it can
be shown that the dominant term in Eq.~(\ref{finerr}) is:
\begin{equation}
\delta\langle u^2\rangle \propto k_c^{3-\nu} \propto N^{\frac{\nu}{3}-1}. 
\end{equation} 
According to this theory, one must determine the exponent $\nu$ of the
singular part of the Green's function $A(k)$, at $k=0$.

For a harmonic lattice:
\begin{equation}
A_{\rm harmonic}({\bf k},\omega) =
\frac{2\pi\hbar}{2m\omega_k}[\delta(\omega-\omega_k) -
\delta(\omega+\omega_k)],
\end{equation}
where $\omega_k$ is the frequency of a phonon of wavevector $k$ and
$m$ is the mass.  In the limit of small k, the dispersion $\omega_k$ is
linear in $k$.  Integrating over $\omega$, we obtain
\begin{equation}
A(k)= \frac{\hbar}{m}\frac{1}{\omega_k}\coth(\frac{\beta\hbar\omega_k}{2})
\label{harmonicak}
\end{equation} 
Thus for $k_BT\gg\hbar\omega_k$ we find $\nu=2$, but at $T\approx 0$,
$\nu=1$, as did Young\cite{young} and Runge\cite{runge}.  A crossover
between the two scaling forms occurs at $\hbar\omega_k \approx k_BT$.

Solid helium is known to have a significant phonon linewidth.  For
small $k$, the phonon linewidth can be approximated to be $\gamma_k
\cong 0.2\omega_k$\cite{ets2,seyfert}.  For a damped harmonic lattice,
\begin{equation}
A_{\rm damped}(k,\omega) =
\frac{\hbar}{2m\omega_k}\left(\frac{\gamma_k}{(\omega-\omega_k)^2+\gamma_k^2/4}
- \frac{\gamma_k}{(\omega+\omega_k)^2+\gamma_k^2/4}\right).
\label{damped_akw}
\end{equation}
Integrating over $\omega$, we find that for $T=0$,
\begin{equation}
A(k) =
\frac{2\hbar}{m\omega_k}\tan^{-1}\left(\frac{2\omega_k}{\gamma_k}\right)
\propto \frac{1}{\omega_k},
\end{equation}
which gives $\nu=1$.  For $T>0$, Eq.~(\ref{damped_akw}) can be
integrated numerically.  As before, we find $\nu=2$ when
$k_BT\gg\hbar\omega_k$ and $\nu=1$ when $k_BT\ll\hbar\omega_k$, with a
crossover at $\hbar\omega_k \approx k_BT$, with width in $\hbar\omega_k/k_BT$ equal to that of the undamped harmonic lattice.  

The width of the crossover region is defined in terms of the ratio
$\hbar\omega_k/k_BT$ (see Fig.  [\ref{crossover}]).  The phonons
excluded from the finite system have linear dispersion with an upper
bound of $\omega_k = sk_c = 2\pi s \rho^{1/3}/N^{1/3}$, where $s$ is
the speed of sound.  By choosing typical values for temperature,
density and sound speed, we can estimate the width of the crossover
region as a function of the number of particles $N$.  For example, if
we assume the speed of sound is $s \cong 4.0 \times 10^{13}$~\AA/sec,
we find that for temperature $T=20.0$~K and density
$\rho=0.055$~\AA$^{-3}$, $N_{{\rm crossover}} \sim 10^2-10^5$.  A
simple power law fit to system sizes in this range is inaccurate.

If we assume a harmonic spectrum, we can use Eq.~(\ref{harmonicak})
to write
\begin{align}
\delta \langle u^2\rangle & \propto \int\limits_0^{k_c}dk\;k^2\;A(k)
\propto
\int\limits_0^{k_c}dk\;k\coth(\frac{\beta\hbar\omega_k}{2})\nonumber
\\ & \propto N^{-1/3}\left( 1 +
\log\left(\frac{1-\exp(-BN^{-1/3})}{BN^{-1/3}}\right)\right) +
{\cal O}(N^{-1}),
\end{align}
where $B\equiv 2\pi\beta\hbar s \rho^{1/3}$.  The functional form of
\usq can now be written as
\begin{equation}
\langle u^2\rangle = \langle u^2\rangle_\infty - AN^{-1/3}\left( 1 +
\log\left(\frac{1-\exp(-BN^{-1/3})}{BN^{-1/3}}\right)\right).
\label{usqinfform}
\end{equation}
The parameter $B$ is not effectively constrained by the PIMC values of
$\langle u^2\rangle$ versus $N$, and therefore cannot be accurately
determined by fitting.  Instead, by using physically reasonable values
of $B$, we can use least-squares fitting (of $A$ and $\langle
u^2\rangle_\infty$) to extrapolate our PIMC data to the thermodynamic
limit and get values for $\langle u^2\rangle_\infty$ (see Table
\ref{usqinftable}.)  The speed of sound $s$ can be obtained from
experimental measurements of isothermal compressibility
$\kappa_T$\cite{driessen}, using the relationship
\begin{equation}
s^2 = \frac{1}{\rho m \kappa_T},
\end{equation}
However, we obtained poor fits using the value of $B$ obtained in this
manner.  The lack of self-consistency with physical parameters shows
that the undamped harmonic lattice used to derive
Eq. (\ref{usqinfform}) is insufficient for describing the mean-squared
displacement in solid helium.  A larger range of $B$ was required for
reasonable fitting, both due to anisotropy in $s$ and the
approximations used in calculating the functional form
Eq.~(\ref{usqinfform}).  We were able to obtain reasonable fitting by
using values of $B$ which correspond to $s = 4000~\pm~2000$~m/sec,
which is larger than experimental values by about a factor of 2 on
average.  For systems away from melting, the fitting was insensitive
to the value of $B$, and the propagated error from $B$ was on the same
order of magnitude as the fitting error, in estimating $\langle
u^2\rangle_\infty$.  The reduced $\chi^2$ for systems away from
melting was typically between 0.1 and 1.0.  For systems near melting,
the fitting was much more sensitive to the estimated range of $B$,
particularly the lower bound, and the errorbar on \usqinf was
completely determined by the uncertainty in $B$.  The reduced $\chi^2$
for these systems was usually between 5 and 10.  A more accurate
functional form, perhaps one which takes the phonon linewidth into
account, would greatly improve the accuracy of extrapolating to the
thermodynamic limit for these systems.

Our PIMC calculations agree with all direct scattering measurements of
$\langle u^2\rangle$, when extrapolated to the thermodynamic limit.  We are able to
confirm both computational and experimental methods to an accuracy of
5\% in the mean-squared displacement.  Near melting, the accuracy of
the PIMC values was considerably reduced, due to uncertainty in the
functional form of the finite-size effects.  PIMC values were
generally higher than the indirect scattering measurements\cite{ets1}.
Because the indirect measurements assumed contributions from
single-phonon processes only, this discrepancy gives evidence for the
importance of multi-phonon processes in solid helium.

\section{Temperature Dependence}

Using Eq.~(\ref{harmonicak}), it is straightforward to show that the
harmonic approximation predicts \usqinf to have temperature dependence
of the form
\begin{equation}
\langle u^2\rangle_\infty  = \langle u^2\rangle^{T=0}_\infty + C\,T^2.
\end{equation}
However, our extrapolated values of \usq appear to fit to a $T^3$
power law, rather than $T^2$ for $5~{\rm K}~\le~T~\le 20~{\rm K}$ (see
Fig.~[\ref{usqinftempdep}]).  With less than a decade in temperature,
it is difficult to draw any direct conclusions from this, although it
may indicate that harmonic theory is insufficient to accurately
describe solid helium, and that higher order terms dominate the
temperature dependence.

\section{Non-Gaussian Corrections to \usq}

We have determined the deviation of the density from a Gaussian
distribution by two methods.  The first was fitting $\ln(S(k))$ to a
polynomial in $k^2$.  As shown in Eq.~(\ref{sk5}), the linear term is
\usq, the quadratic term is proportional to the kurtosis $\kappa$. We
also directly calculated the kurtosis in the $\langle 111\rangle$ and
$\langle 100 \rangle$ directions, using Eq.~(\ref{kurt1}). The kurtosis was found to be
non-zero and anisotropic in the fcc solid helium systems we studied.

Shown are graphs with kurtosis in the $\langle 100 \rangle$ direction
as a function of density and temperature. The kurtosis is roughly
twice as large in systems near melting.  Away from melting, the
kurtosis appears to be independent of both temperature and density.

We found that $\kappa$ is independent of $N$, for values of $N \ge
500$.  For smaller system sizes, the finite-size effects are large,
but drop off quickly with increasing $N$.  However, at the wavevectors
$k \le 9$~\AA$^{-1}$, the effect of the kurtosis is only a few
percent, making this term difficult to observe in scattering
experiments\cite{seyfert,arms,chitra,stassis,burns}.  Vitiello et al
\cite{vitiello} have computed kurtoses of 0.051 and 0.042 at zero
temperature using Shadow wave functions, for molar volumes of 20.5 and
18.3.  These values are consistent with the values computed here.

\section{Conclusions}

PIMC simulations of the Debye-Waller factor in solid helium agree with
experimental results to better than 5\% accuracy, indicating that the
assumed potential, the computational methods, and the experimental
analysis are correct within the stated errors.  We determined the
first non-Gaussian contribution, a directionally-dependent kurtosis.
The finite-size effects were found to be in a crossover region between
the classical and quantum scaling limits, and hence a power-law
dependence was insufficient for extrapolation to the thermodynamic
limit.  The harmonic approximation gave an approximate functional form
for $\langle u^2\rangle$ which was used to extrapolate finite values
to the thermodynamic limit.  The extrapolated values agree with all
available direct scattering measurements, although the extrapolation
became more sensitive with increased proximity to the melting
transition, with correspondingly larger uncertainties.  Higher
extrapolation accuracy could be obtained with a more accurate scaling
form.  The effective temperature dependence of $\langle u^2\rangle$
appears to be closer to $T^3$, rather than the $T^2$ predicted by
harmonic theory.

\begin{acknowledgements}
The authors wish to thank R. O. Simmons, G. Baym, G. H. Bauer, and
N. Goldenfeld, for useful discussions.  This research was carried out
on the Origin 2000 at the National Center for Supercomputing
Applications, and was supported by the NASA Microgravity Research
Division, Fundamental Physics Program.
\end{acknowledgements}

\begin{table}[hbt]
\caption{PIMC results.  \usqinf was estimated by fitting finite PIMC
data to the function given in Eq.~(\ref{usqinfform}), assuming a sound
speed $s = 4000 \pm 2000$~m/sec.  The listed uncertainty in \usqinf
represents the range of fitted values corresponding to the uncertainty
in $s$.  The errors in the energies represent the statistical
uncertainties in the final digit, while the experimental errors are
the total uncertainties as estimated by the authors.  $\kappa_{100}$
represents the estimated kurtosis in the $\langle 100\rangle$
direction.  $\kappa$ was estimated for the largest system size
available, usually $N=1372$.}

\begin{tabular}{ldddddddd}
type & $V_m ({\rm cm}^3)$ & T~(K) & $T_{{\rm expt}}$~(K) & $\langle u^2\rangle_\infty (10^{-2}$~\AA$^2)$ & $\langle
u^2\rangle_{{\rm expt}} (10^{-2}$~\AA$^2)$ & $\kappa_{100}$ & $E_{{\rm
kinetic}}$~(K) & $E_{{\rm tot}}$~(K)
\\
\tableline
 fcc $^4$He & 10.98 & 20.00 & 20.25 & 9.77(39) & 9.99(27)\tablenotemark[1]& 0.09 & 80.30(2) & 50.24(2)\\
 fcc $^4$He & 10.98 & 17.78 & & 9.15(32) & & 0.08 & 79.10(4) & 48.02(5) \\
 fcc $^4$He & 10.98 & 16.84 & & 8.71(23) & & 0.07 & 79.15(6) & 47.52(6)\\
 fcc $^4$He & 10.98 & 16.00 & & 8.76(21) & & 0.07 & 78.51(4) & 46.64(4) \\
 fcc $^4$He & 10.98 & 15.24 & & 8.49(18) & & 0.06 & 78.61(6) & 46.44(6) \\
 fcc $^4$He & 10.98 & 13.33 & & 8.36(16) & & 0.06 & 77.63(4) & 45.08(4) \\
 fcc $^4$He & 10.98 & 10.00 & & 8.02(8) & & 0.06 & 77.06(2) & 43.99(2) \\
 fcc $^4$He & 10.98 & 8.00  & & 7.89(7) & & 0.06 & 76.87(4) & 43.67(3) \\
 fcc $^4$He & 10.98 & 5.00  & & 7.79(5) & & 0.06 & 76.80(2) & 43.50(2) \\
 fcc $^4$He & 10.39 & 24.60 & 24.40 & 8.89(53) & 8.23(40)\tablenotemark[2] & 0.09 & 90.41(4) & 68.71(5) \\
 fcc $^4$He & 10.02 & 26.67 & 25.94 & 8.29(61) & 8.43(18)\tablenotemark[2] & 0.09 & 96.48(4) & 81.53(4) \\
 fcc $^4$He & 9.02 & 35.56 & 38.00 & 6.91(94) & 5.66\tablenotemark[3] & 0.09 & 118.14(7) & 133.74(8) \\
 fcc $^4$He & 9.02 & 22.86 & & 5.77(21) & & 0.04 & 110.68(5) & 120.77(5)\\
 fcc $^4$He & 9.43 & 21.33 & & 6.31(21) & & 0.05 & 109.14(6) & 114.16(6) \\
 fcc $^4$He & 9.97 & 26.67 & 28.00 & 8.14(58) & 6.93\tablenotemark[3] & 0.08 & 101.33(5) & 94.26(5) \\
 fcc $^4$He & 9.97 & 18.82 & 19.00 & 7.04(20) & 6.20\tablenotemark[3] & 0.05 & 98.82(6) & 88.03(6) \\
\tableline
 fcc $^3$He & 11.54 & 17.78 & 18.13 & 11.41(40) & 11.43(11)\tablenotemark[1]& 0.10 & 86.14(3) & 58.53(3) \\
 fcc $^3$He & 11.54 & 10.00 & & 10.00(12) & & 0.08 & 84.15(4) & 54.68(4) \\
 fcc $^3$He & 11.54 & 5.00 & & 9.72(5) & & 0.12 & 83.88(5) & 54.22(4) \\
 fcc $^3$He & 10.98 & 17.78 & & 9.87(28) & & 0.09 & 93.18(5) & 70.39(5) \\
 fcc $^3$He & 10.00 & 29.09 & & 9.07(76) & & 0.09 & 113.74(5) & 110.00(6)\\
\tableline
 hcp $^4$He & 12.12 & 14.55 & 14.23 & 11.17(13) & 11.25(28)\tablenotemark[1] & & 66.44(4) & 27.95(4) \\
 hcp $^4$He & 12.12 & 11.85 & 12.00 & 10.24(13) & 10.26(17)\tablenotemark[1] & & 65.56(6) & 26.16(6) \\
 hcp $^4$He & 12.12 & 5.00 & & 9.52(5) & & & 64.37(3) & 24.15(3) \\
 hcp $^4$He & 15.72 & 5.71 & 5.80 & 17.40(13)\tablenotemark[5] & 17.31(18)\tablenotemark[4] & & 40.66(3) & 0.78(3) \\
 hcp $^4$He & 10.98 & 5.00 & & 7.78(3) & & & 76.74(4) & 43.47(3) \\
\tableline
 hcp $^3$He & 11.90 & 16.84 & 16.81 & 11.84(24) & 11.96(26)\tablenotemark[1] & & 82.17(6) & 52.13(6) \\
 hcp $^3$He & 12.81 & 12.31 & 12.54 & 13.26(19) & 13.43(27)\tablenotemark[1] & & 71.14(4) & 36.37(4)

\end{tabular}
\label{usqinftable} \tablenotetext[1]{Arms and Simmons\cite{arms}.
Direct x-ray measurements.} \tablenotetext[2]{Venkataraman and Simmons\cite{chitra}.  Direct x-ray measurements.}
\tablenotetext[3]{Thomlinson, Eckert and Shirane\cite{ets1}.  Indirect
neutron measurements .}  \tablenotetext[4]{Stassis, Khatamian, and
Kline\cite{stassis}.  Direct neutron measurements.}
\tablenotetext[5]{Extrapolated using speed of sound $s = 1000 \pm 200
$~m/sec, based on experimental data near this density\cite{brun}.}
\end{table}

\begin{figure}[hbt]
\caption{Numerical calculation of $A(k)$ as a function of
$\hbar\omega_k/k_BT$, for a damped harmonic oscillator with linewidth
$\gamma_k = 0.2\omega_k$.  Shown is the crossover between the quantum
limit (dashed line) and the classical limit (dotted line), which
allows one to estimate the range of system sizes affected by the
crossover region.}
\label{crossover}
\end{figure}

\begin{figure}[hbt]
\caption{Extrapolation of finite PIMC simulation data to the
thermodynamic limit, for fcc \he, $V_m=10.98$~cm$^3$, T=20.0 K.  \usq
was calculated by direct averaging (open circles) and by fitting to
$S(k)$ data (open triangles).  Both methods always agreed within
statistical error bars.  Least-squares fitting was used to fit the
directly averaged data to Eq.~(\ref{usqinfform}).  The experimental
data point (solid circle) is from Arms and Simmons.}
\label{bdw50}
\end{figure}

\begin{figure}[hbt]
\caption{Temperature dependence of \usqinf, using the extrapolated
values from Table \ref{usqinftable}.  Shown are values for fcc $^4$He,
$V_m=10.98$~cm$^3$ (solid circles), and fcc $^3$He, $V_m=11.54$~cm$^3$
(solid squares).}
\label{usqinftempdep}
\end{figure}

\begin{figure}[hbt]
\caption{$\frac{1}{k^2}\ln(\frac{S(k)-1}{N-1})$ versus $k^2$ for fcc
\he, $V_m=10.98\; cm^3$, N=864, T=20.0 K.  The kurtosis $\kappa$ is
given by the slope, and agrees with direct calculations of
$\langle u^4\rangle/\langle u^2\rangle^2-3$.  In the $\langle
100\rangle$ direction (dashed line), the fitted kurtosis was
$0.11\pm0.03$, while the direct value was $0.09\pm0.03$. In the
$\langle 111\rangle$ direction (dotted line), the fitted kurtosis was
$0.05\pm0.03$, while the direct value was $0.02\pm0.03$.}
\label{vvskdir}
\end{figure}

\begin{figure}[hbt]
\caption{Kurtosis $\kappa$ vs. molar volume $V_m$ and temperature $T$,
in the $\langle 100 \rangle$ direction, for fcc $^4He$.  The kurtosis
is noticeably larger near the experimental melting line, but is
otherwise independent of temperature and density.}
\label{kurtvol}
\end{figure}

\pagebreak

\begin{center}
\epsfig{file=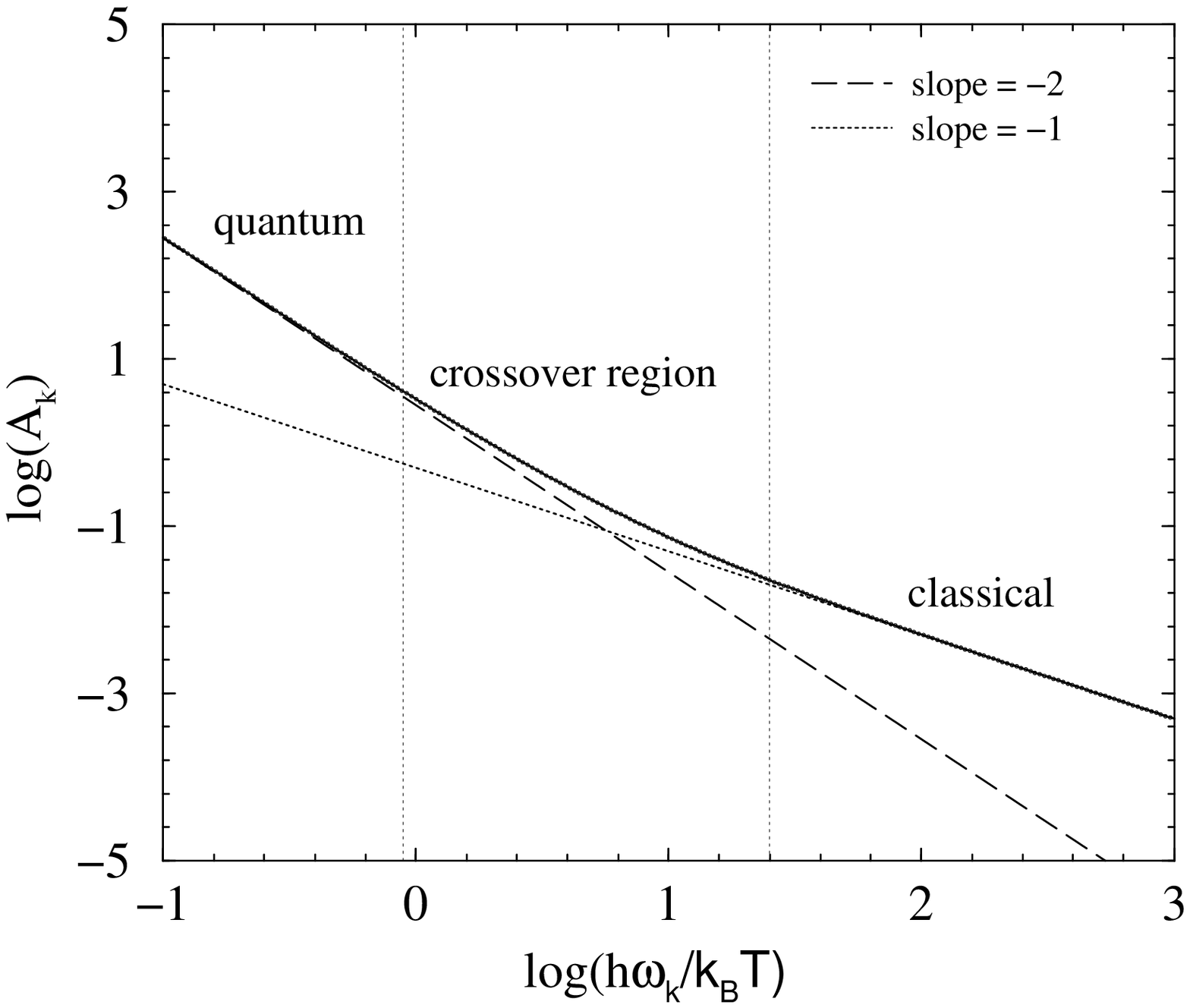}
\end{center}
\vfill
Draeger Figure 1

\begin{center}
\epsfig{file=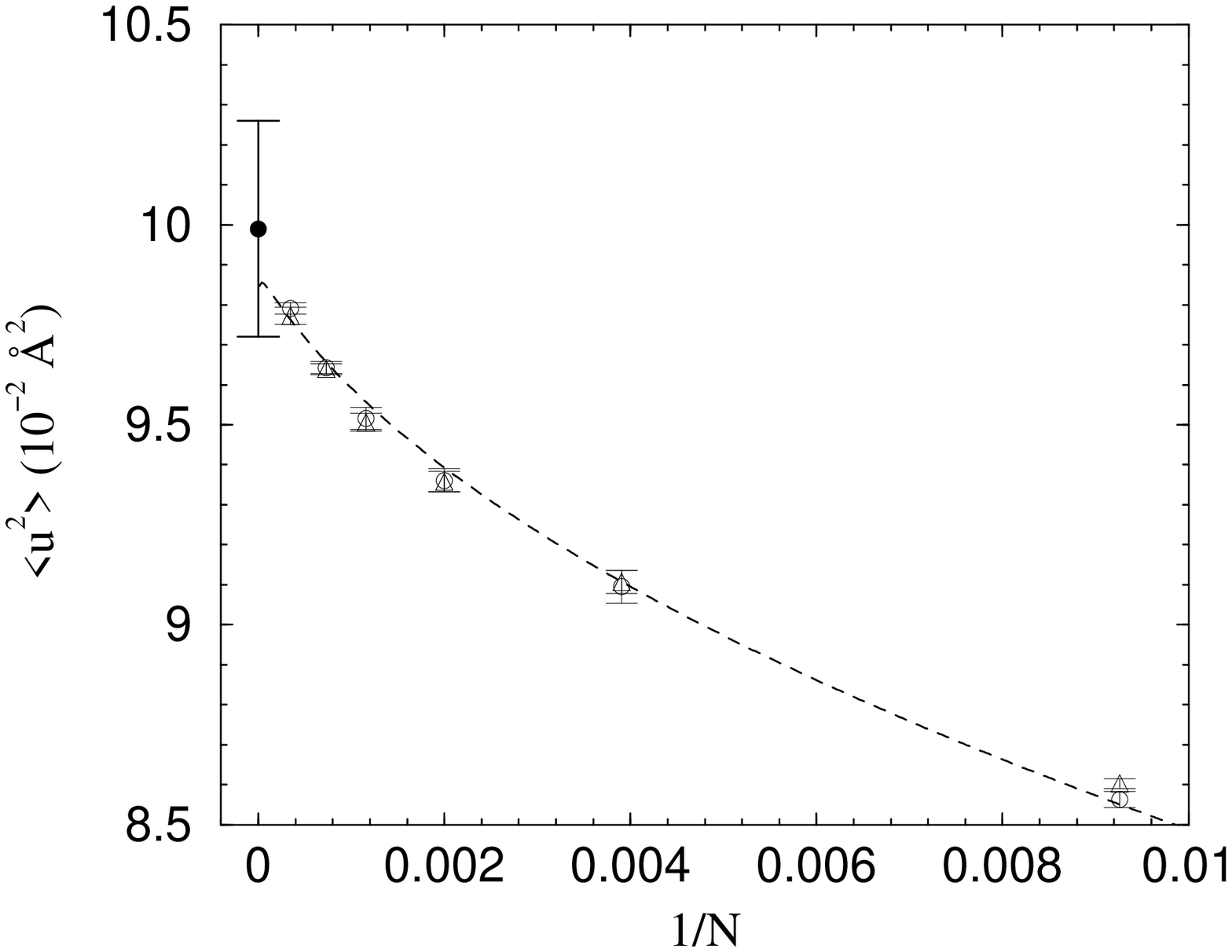}
\end{center}
\vfill
Draeger Figure 2

\begin{center}
\epsfig{file=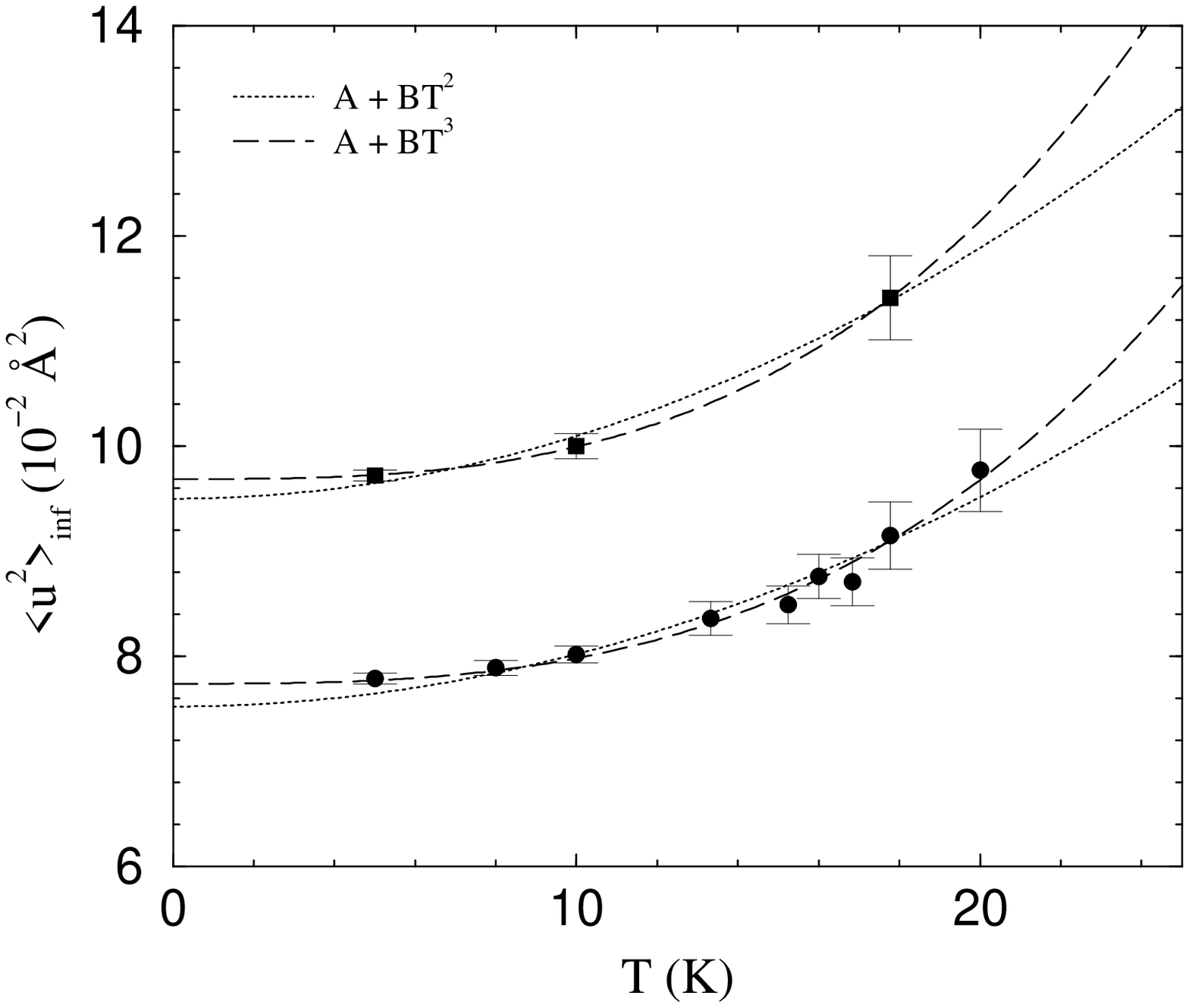}
\end{center}
\vfill
Draeger Figure 3

\begin{center}
\epsfig{file=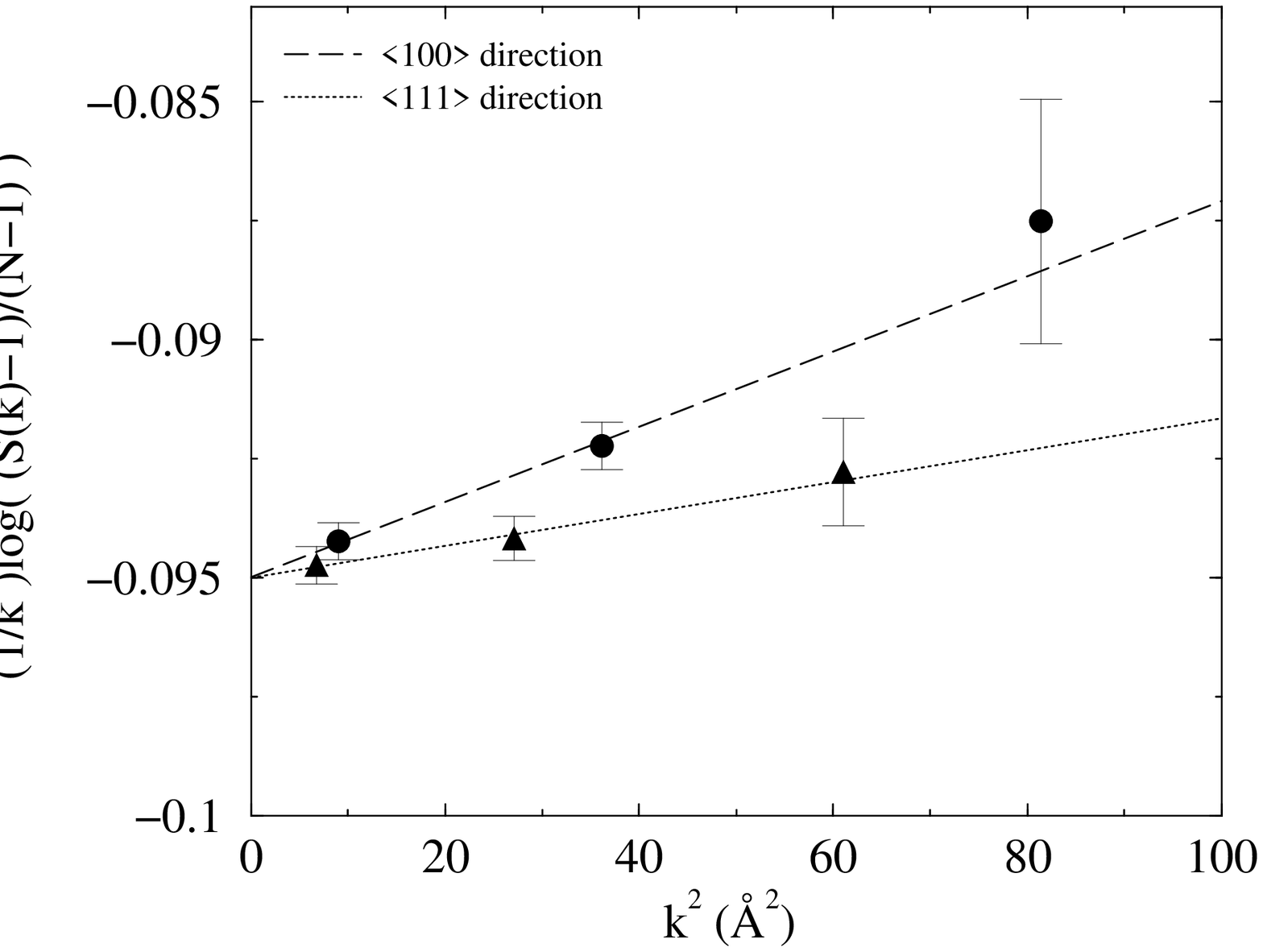}
\end{center}
\vfill
Draeger Figure 4

\begin{center}
\centerline{\epsfig{file=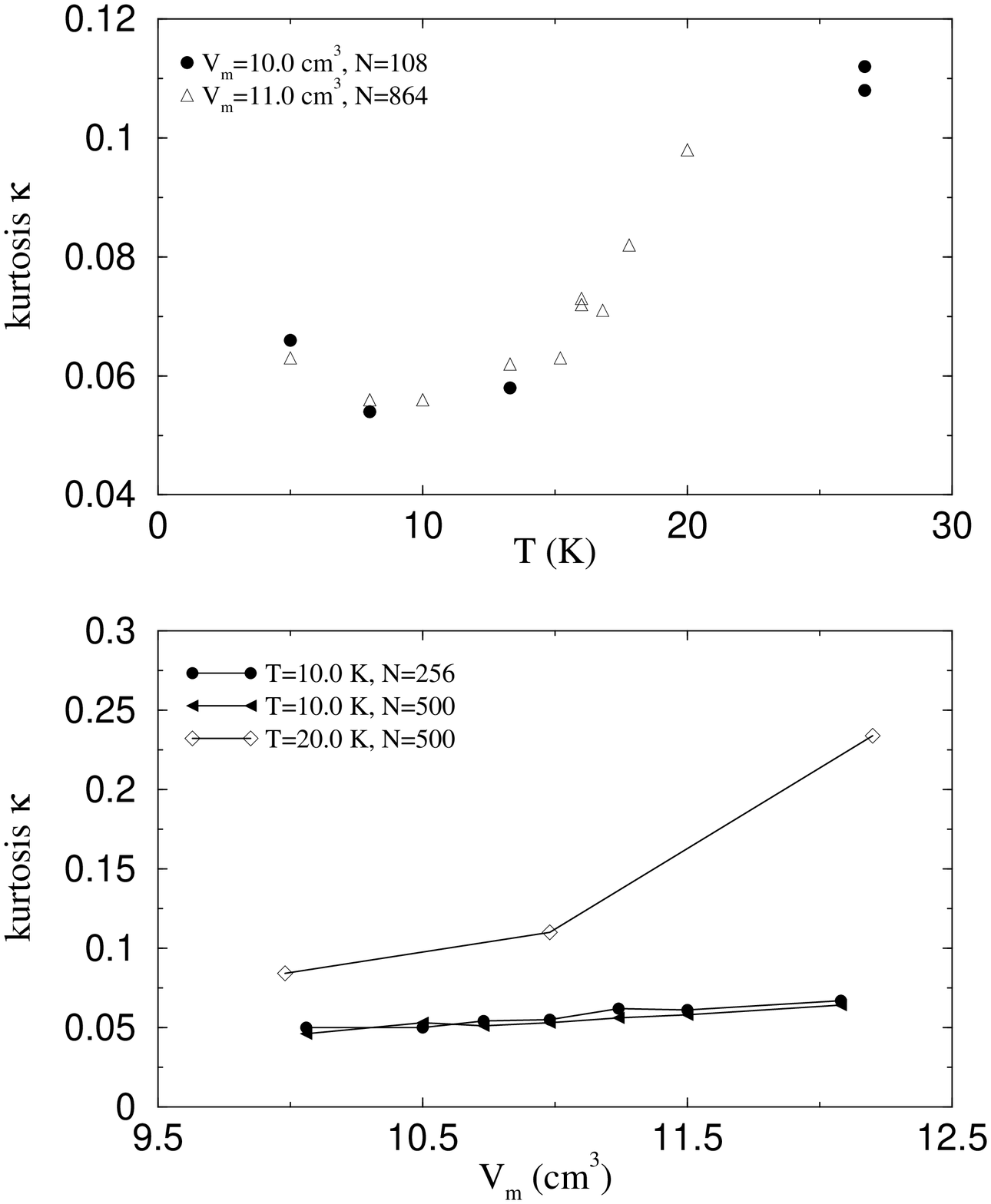}}
\end{center}
\vfill
Draeger Figure 5

\end{document}